\documentclass[aps,preprint,nofootinbib,preprintnumbers,eqsecnum,12pt]{revtex4-1}
\usepackage[usenames, dvipsnames]{color}

\usepackage{graphicx}
\usepackage{xcolor}
\usepackage{rotating}
\usepackage{bm,amsmath,amssymb}
\usepackage[utf8]{inputenc}
\usepackage{footmisc}

\newcommand{\nb}[1]{\color{blue}}

\newcommand{\ML}[1]{{\textcolor{magenta}{#1}}}
\newcommand{\hl}[1]{\color{magenta}}

\usepackage{yfonts}

\usepackage[
	  pagebackref=false,
	  colorlinks=true,
      linkcolor=blue,
      urlcolor=blue,
      filecolor=black,
      citecolor=red,
      pdfstartview=FitV,
      pdftitle={},
        pdfauthor={},
        pdfsubject={},
        pdfkeywords={},
        pdfpagemode=None,
        bookmarksopen=true
      ]{hyperref}

\usepackage{graphicx}
\graphicspath{ {./images/} }

\usepackage{caption}
\usepackage{subcaption}

\usepackage{comment}

\usepackage[normalem]{ulem}
\usepackage{amsmath}
\usepackage{enumerate}
\usepackage{amsfonts}
\usepackage{epsfig}
\usepackage{mathbbol}

\def\Im{\mathop{\rm Im} }

\newcommand\half{{\ensuremath{\frac{1}{2}}}}
\newcommand\p{\ensuremath{\partial}}
\newcommand\pp{\ensuremath{\vec\nabla}}

\newcommand{\be}{\begin{equation}}
\newcommand{\ee}{\end{equation}}
\newcommand{\bea}{\begin{eqnarray}}
\newcommand{\eea}{\end{eqnarray}}
\newcommand{\bega}{\begin{gather}}
\newcommand{\eega}{\end{gather}}

\newcommand{\bi}{\begin{itemize}}
\newcommand{\ei}{\end{itemize}}
\newcommand{\ben}{\begin{enumerate}}
\newcommand{\een}{\end{enumerate}}
\newcommand{\bca}{\begin{cases}}
\newcommand{\eca}{\end{cases}}
\newcommand{\bln}{\begin{align}}
\newcommand{\eln}{\end{align}}
\newcommand{\bst}{\begin{split}}
\newcommand{\est}{\end{split}}
\def\ie{\begin{equation}\begin{aligned}}
\def\fe{\end{aligned}\end{equation}}
\newcommand{\bma}{\le(\begin{matrix}}
\newcommand{\ema}{\end{matrix}\ri)}
\newcommand{\bwt}{\begin{widetext}}
\newcommand{\ewt}{\end{widetext}}

\newcommand\lam{\lambda}

\newcommand\Th{{\Theta}}

\newcommand\ov{\over}
\newcommand\ha{{\half}}

\def\le{\left}
\def\ri{\right}

\newcommand\sA{{\ensuremath{{\mathcal A}}}}

\newcommand\sC{{\ensuremath{{\mathcal C}}}}

\newcommand\sF{{\ensuremath{{\mathcal F}}}}

\newcommand\sL{{\ensuremath{{\mathcal L}}}}

\newcommand\sP{{\ensuremath{{\mathcal P}}}}

\newcommand\sT{{\mathcal T}}

\DeclareMathAlphabet{\pazocal}{OMS}{zplm}{m}{n}

\begin{document}

\preprint{}

\title{Active actions: effective field theory for active nematics}

\author{Michael J. Landry}
\affiliation{Center for Theoretical Physics, MIT, Cambridge, MA 02139, USA }

\begin{abstract}

Active matter consumes energy from the environment and transforms it into mechanical work. Notable examples from biology include cell division, bacterial swarms, and muscle contraction. In this work, we investigate the nature of active matter systems using the powerful effective field theory toolbox. This allows us to construct the most general theory without ambiguity up to a given order in the derivative expansion. Our primary focus is active nematics---liquid crystal systems that spontaneously break rotational but not translational symmetry---in two spatial dimensions. (Such spontaneous symmetry breaking is allowed if the nematic is embedded in a higher dimensional space.) While we focus on this one particular class of physical system, the tools developed here can in principle be applied to any active matter system. Our theories give unambiguous predictions for the relationship between fluctuations and equations of motion in the presence of activity, generalizing the standard fluctuation-dissipation relations characteristic of passive systems. 

\end{abstract}

\maketitle
\tableofcontents

\section{Introduction} 

Active matter is a broad area of study that encompasses biological and synthetic systems that exist far from thermodynamic equilibrium~\cite{RevModPhys.85.1143}. Such systems consume energy and convert it into mechanical work, which allows for self-propulsion. In this way, all biological systems fall under the active matter umbrella~\cite{needleman2017active}. The nature of this energy source can vary from system to system. In many biological examples, energy is stored in chemical bonds and is then released to perform work---this is how we are able to move the muscles in our bodies. But the energy source could be something entirely different, like a light that shines on an active sample~\cite{RevModPhys.88.045006}.

Apart from being the fundamental physical principle that undergirds all of life, active matter systems offer an intriguing space for exploring non-equilibrium physics in a broader context~\cite{Ramaswamy_2010}. There are various theoretical questions that accompany the study of active systems. When constructing a theory of a passive system, principles of local equilibrium can be invoked that constrain both the equations of motion and the statistical fluctuations~\cite{RKubo_1966}. When studying active systems, however, it is not so clear how principles of local equilibrium are relevant. On the one hand, active systems are very far from equilibrium---they continually consume energy and use it to move. %\footnote{We emphasize that there are many far-from-equilibrium systems that consume energy---e.g. steady-state flows---that are not active. Active systems must be able to convert the energy they consume into motion.} 
On the other hand, notions like temperature or chemical potential---hydrodynamic quantities that are only well-defined in equilibrium~\cite{Kovtun:2012rj}---are often relevant. 
So while in the most general case in which we are totally agnostic about the origin of activity, no notion of local equilibrium can be universally defined, in many real-world applications, some notion of local equilibrium must persist. %A major aim of this paper is to understand the meaning and significance of local equilibrium in active systems.

The particular system we will focus on is active nematics in two spatial dimensions. These are apolar liquid crystal systems consisting of elongated molecules that spontaneously break spatial rotation symmetry~\cite{gennes1993the}. They consume energy from their surroundings and convert it into mechanical work~\cite{simha2002hydrodynamic,Doostmohammadi1}. 
Active two-dimensional nematic order has been identified in numerous biological systems. These systems encompass a variety of phenomena, such as epithelial monolayers~\cite{saw2017topological,kawaguchi2017topological,blanch2018turbulent} and suspensions of cytoskeletal filaments~\cite{sanchez2012spontaneous,keber2014topology}. On the theoretical front, much attention has been paid to topological defects~\cite{vafa2020multi-defect,vafa2022defectAbsorption,vafa2022defectDynamics,vafa2023active,vafa2022active} and resulting nematic ``turbulence,'' which occurs when the complex dynamics of topological defects are driven by strong active elastic distortions of the nematic substrate~\cite{tan2019topological,alert2022active,PhysRevLett.110.228101,ellis2018curvature,PhysRevX.9.041047,putzig2016instabilities,srivastava2016negative,doostmohammadi2016stabilization,Pearce_2021,Thijssen_2020,thampi2016activeTurbulence,giomi2015geometry}.

The aim of this paper is to extend the effective field theory (EFT) paradigm to account for activity, focusing on nematics in two dimensions. In the passive case, techniques for constructing actions on the Schwinger-Keldysh (SK) contour that account for the non-equilibrium dynamics of many body systems---including nematics---have been well-studied~\cite{Landry:2019iel,Baggioli:2020haa,Landry:2020ire}. An important element of passive dynamics is the dynamical KMS symmetry~\cite{Crossley:2015evo,Glorioso:2016gsa,Glorioso:2017fpd,Glorioso:2018wxw}. This is a symmetry of the action that implements both local thermal equilibrium and microscopic time-reversibility. It ensures that the correlation functions of observables satisfy the KMS conditions, which are real-time constraints that define thermal equilibrium. The primary obstacle to constructing active EFTs is that dynamical KMS symmetry appears to automatically ensure passive dynamics. So how can any notion of local equilibrium be implemented if activity is present? To answer this question we we will investigate two approaches: 
\begin{enumerate}
     \item The most straight-forward approach is to maintain the usual dynamical KMS symmetry, but couple the passive system to a far-from-equilibrium sector. For example, we could introduce a $U(1)$ charge that very slowly decays. This might represent e.g. a slow chemical reaction like the consumption of ATP that drives activity in a cell. Such an approach gives a clear account of the origin of activity at the price of being model-dependent. 
     \item To construct a more model-independent active EFT, we would like to modify the dynamical KMS relations in such a way that active dynamics emerge. Note that we could simply do away with dynamical KMS symmetry entirely, but this would remove all notion of local equilibrium, which as discussed previously, is undesirable. We will therefore take inspiration from the above approach and propose such a modified dynamical KMS~symmetry.
\end{enumerate}

The paper is organized as follows. In Section~\ref{passive} we review how to construct non-equilibrium EFTs on the SK contour for passive nematics. Working first in the probe limit, we show how the EFT can be modified to include passive velocities in open systems. Next, in Section~\ref{active dyn} we couple the passive action to a far-from-equilibrium sector that drives activity in the system. We find a leading-order correction to the usual active constitutive relation for the velocity. Then, taking inspiration from this EFT, we propose a modification to dynamical KMS symmetry in Section~\ref{mod dKMS}.
And finally in Section~\ref{discussion} we discuss the implications and generalizations for such active EFTs.

\section{Passive dynamics}\label{passive}

Before investigating the active case, we begin by considering passive dynamics for nematics in two spatial dimensions. As a particularly simple example, we begin with the probe limit in which the only degree of freedom is the Goldstone corresponding to spontaneously broken rotations.\footnote{In two dimensions, passive nematic phase should be impossible as SSB cannot occur. The correct way to address this issue is account for the embedding of the two dimensional nematic into three dimensions. For simplicity, however, we will merely circumvent this issue by refraining from computing loop diagrams.} Unfortunately, such an EFT has no notion of local velocity, so endowing it with activity directly would be tricky. Moreover, introducing local velocity is typically accomplished by considering the conservation of the stress-energy tensor. But in active systems, energy and momentum can be exchanged with the environment, so it ought not be conserved. To fix these problems, we propose a way to introduce dynamical velocity without conserving the stress-energy tensor, which will ultimately make including activity quite easy.

\subsection{Probe limit}

A nematic liquid crystal consists of oblong molecules that spontaneously break spatial rotation symmetry, but leave space and time translations unbroken. As a result, in the infrared, the only relevant degrees of freedom are the rotation Goldstone $\theta$ and its $a$-type partner $\theta_a$. It is convenient to define the vector pointing parallel to the oblong molecules by $\vec n = (\cos\theta,\sin\theta)$. In nematic phase, flipping the oblong molecules by 180 degrees leaves the system unchanged, so we postulate the $Z_2$ symmetry $\vec n\to -\vec n$. It is sometimes convenient to work with the symmetric, traceless tensor $Q^{ij} = n^i n^j - \ha \delta^{ij}$, which is invariant under this $Z_2$ symmetry. 

Suppose the system exists at fixed finite temperature $T_0=1/\beta_0$. The effective action $I_{\rm EFT}[\theta,\theta_a]$ for the rotation Goldstones can be constructed using the framework of~~\cite{Crossley:2015evo,Glorioso:2016gsa,Glorioso:2017fpd,Glorioso:2018wxw,Landry:2019iel}. In particular, it must satisfy the following conditions:
\begin{enumerate}
     \item It enjoys translational invariance in $x^\mu=(t,x^i)$ and rotational invariance on $x^i$ coordinates. The latter symmetry acts non-trivially on the Goldstone by
     \be
          x^i\to R^{ij}(\phi) x^j,\quad \theta\to\theta+\phi,\quad R(\phi)\in SO(2). 
     \ee
     \item There are various unitarity constraints given by
     \bega\label{unitary1}
          I^*_{\rm EFT}[\theta,\theta_a] = - I_{\rm EFT}[\theta,-\theta_a],\\ \label{unitary2}
          \Im \, I_{\rm EFT} \geq 0 ,\\ \label{unitary3}
          I_{\rm EFT}[\theta,\theta_a=0] = 0.  
     \end{gather}
     These constraints will be straightforwardly generalized when other fields are included. 
     \item It is invariant under dynamical KMS transformations, that is $I_{\rm EFT}[\theta,\theta_a] = I_{\rm EFT}[\tilde \theta,\tilde\theta_a]$, where dynamical KMS symmetry is a $Z_2$ transformation of the form
     \be
          \tilde \theta = \Th \theta,\quad \tilde \theta_a = \Th \theta_a + i \beta_0 \Th \p_t \theta. 
     \ee
     Here $\Th$ is a discrete anti-unitary, time-reversing symmetry of the microscopic theory. For example it could be $\Th=\sT$ or $\Th=\sC\sP\sT$. Dynamical KMS symmetry in conjunction with the unitarity constraints~\eqref{unitary1}-\eqref{unitary3} can be used to deduce an entropy current $s^\mu$ with non-negative divegence $\p_\mu s^\mu\geq 0$. That is, the second law of thermodynamics follows automatically~\cite{Glorioso:2016gsa}.
     \item Impose whatever other discrete symmetries are formed from charge conjugation $\sC$ and parity $\sP$ that are preserved by the underlying microscopic system. For the purposes of this paper, $\sC$ symmetry is irrelevant as there are no objects that carry fundamental charge and we will always impose $\sP$~symmetry. 
\end{enumerate}

Effective actions are organized as derivative expansions, so we must assign weights to derivatives and fields. In the probe limit, boosts are explicitly broken, so spatial and temporal derivatives need not be on equal footing. It will turn out that the leading-order dynamics of passive nematics are diffusive, suggesting the weight-assignments $[\p_t]=2[\p_i]=2$. Additionally, as $\theta\in[0,2\pi)$ it ought to have a weight of zero. Finally, as dynamical KMS symmetry relates $\theta_a$ to $\p_t\theta$, we assign $\theta_a$ a weight of two so that this symmetry does not mix terms of differing weights. 

The symmetry-invariant building-blocks are as follows. Notice that while $\theta$ shifts under rotations, $\theta_a$ does not. As a result, $\theta_a$ can appear on its own without derivatives. Next, the temporal derivative $\p_t\theta$ is manifestly rotation-invariant. To take spatial derivatives, define the covariant derivatives
\bega
     \nabla_1 \theta = \vec n\cdot\pp\theta,\quad \nabla_1\theta_a = - \theta_a \vec n\times \pp \theta + \vec n \cdot \pp \theta_a ,\\
     \nabla_2 \theta = \vec n\times\pp\theta,,\quad \nabla_2 \theta_a = \theta_a \vec n\cdot\pp\theta - \vec n \times\pp \theta_a .
\end{gather}
The covariant derivatives' actions on $\theta_a$ are defined in such a way that dynamical KMS symmetry can be easily implemented~\cite{Landry:2019iel,Landry:2021kko}. Lastly, it is often convenient to define $\hat\theta_a=\theta_a + i \beta_0 \p_t \theta$. 

We are now in a position to construct the leading-order action, which includes only weight-4 terms given by
\be\label{probe}
     \sL = {i \gamma \ov \beta_0}\theta_a\hat \theta_a - K_1 \nabla_1\theta \nabla_1\theta_a - K_2 \nabla_2\theta \nabla_2\theta_a. 
\ee
Here $\gamma$ and the Frank coefficients $K_{1,2}$ are constants;~\eqref{unitary2} requires that $\gamma >0$ and requiring a stable free energy implies $K_{1,2}>0$. In the literature it is common to make the so-called one-constant approximation, which means we fix $K\equiv K_1=K_2$~\cite{chaikin2000principles}. The Lagrangian becomes
\be
     \sL_\text{1-const.} = {i \gamma \ov \beta_0}\theta_a\hat \theta_a - K \p_i \theta \p_i \theta_a,
\ee
whose equations of motion yield simple diffusion
\be
     \p_t \theta = D \pp^2 \theta,\quad D = {K\ov\gamma}. 
\ee
Working with $K_1\neq K_2$, the equations of motion become a bit more complicated 
\be\label{eom a}
     \gamma \p_t \theta = (K_1-K_2) \nabla_1\theta \nabla_2\theta + K_1 \pp\cdot(\vec n \, \nabla_1 \theta) - K_2\pp\times(\vec n \, \nabla_2 \theta). 
\ee

\subsection{The non-Stückelberg trick}

The EFT in the previous subsection is sufficient to describe passive dynamics in the probe limit. In this limit, there is no notion of velocity. When we generalize to the active case, however, including a velocity will be crucial. It is the aim of this subsection to propose how to include a velocity even when the system is open, meaning that the stress-energy tensor is not conserved. 

When the stress-energy tensor is conserved, we typically introduce metric sources $g_{\mu\nu},g_{a\mu\nu}$ and corresponding Stückelberg fields $X^\mu(\sigma),X_a^\mu(\sigma)$ defined on the auxiliary coordinates $\sigma^M$. In the classical limit, we can perform a coordinate transformation so that the action is defined on the physical spacetime coordinates $x^\mu=X^\mu$ at the price of promoting $\sigma^M$ to dynamical fields. The Stückelberg fields must appear in the packages 
\be
     G_{a\mu\nu} = g_{a\mu\nu} + \sL_{X_a} g_{\mu\nu},
\ee 
where $\sL_{\xi}$ is the Lie derivative with respect to $\xi$, and all free Lorentz indices must be contracted with $g_{\mu\nu}$. 
Finally, when spacetime translations are unbroken the EFT typically enjoys time-independent diffs 
\be
     \sigma^0\to \sigma^0 + f(\sigma^i),\quad \sigma^i\to \Sigma^i(\sigma^j) .
\ee

In the active case, we will want an EFT that has active velocity, but fixed background temperature $T_0=1/\beta_0$. As a result, we have no need of the temporal Stückelberg fields above, so we fix $\sigma^0=t$ and $X_a^t=0$. The local velocity field is given by
\be
     v^i \equiv {J^i\ov J^t},\quad J^\mu = \epsilon^{\mu\nu\lam\rho} \p_\nu \sigma^1 \p_\lam \sigma^2 \p_\rho \sigma^3. 
\ee 
Next, active systems are open, so the stress-energy tensor cannot be conserved. We thus must suppose that $X_a^i,\sigma^i$ are not true Stückelberg fields. Nevertheless, if the non-conservation of the stress-energy tensor is too severe then the energy given to the system by the fuel-source will be dissipated rapidly and will not lead to activity. Therefore we must suppose that the stress-energy tensor is only weakly non-conserved. As a result, we shall treat $X_a^i,\sigma^i$ as {\it approximate} Stückelberg fields~\cite{Landry:2020tbh,Landry:2021kko}. To this end, decompose the Lagrangian by
\be\label{non stuc}
     \sL = \sL^\text{(non)} + \sL^\text{(Stüc)},
\ee
where $\sL^\text{(non)}$ contains non-Stückelberg terms, e.g. $X_a^i$ without derivatives; while $\sL^\text{(Stüc)}$ contains only Stückelberg terms. As $X_a^i,\sigma^i$ are approximate Stückelberg fields, we suppose that $\sL^\text{(non)}$ is small even when it is lower-order in the derivative expansion. 

Now that the system has a non-trivial velocity, the dynamical KMS transformations must be modified. We have in particular
\be\label{dKMS1}
     \tilde \theta = \Th \theta,\quad \tilde \sigma^i = \Th \sigma^i,\quad \tilde \theta_a = \Th\theta_a + i\beta_0 \Th D_t\theta,\quad \tilde X_a^i = \Th X_a^i + i\beta_0 \Th v^i,
\ee
where $D_t\equiv \p_t + \vec v\cdot \pp$ is the material derivative.  To assign weights to the fields we follow the power-counting scheme of the previous subsection $[\p_t]=2[\p_i]=[\theta_a]=2$ and $[\theta]=0$. To ensure that dynamical KMS symmetry does not mix terms of different weights, it is natural to suppose $[v^i]=[X_a^i]=1$. 

Begin by constructing the non-Stückelberg sector. The leading-order terms are of weight-2, given by
\be
     \sL^\text{(non)} = {i\Lambda^{-1}_{ij} \ov\beta_0} X_a^i \hat X_a^j,\quad \hat X_a^i = X_a^i + i\beta_0 v^i,\quad \Lambda^{ij} = \Lambda_0 \delta^{ij}+ \Lambda_Q Q^{ij}.
\ee
Let us now make use of field redefinitions. We may redefine $X_a^i$ by
\be
     X_a^i \to X_a^i + i f_a^i,
\ee
where $f_a^i$ is an (almost) arbitrary higher-weight vector. Such a field redefinition must however yield an action that is consistent with~\eqref{unitary1}--\eqref{unitary3}. As a result,  $f_a^i$ must be at least linear in $a$-type fields and each occurrence of an $a$-type field must be accompanied by a factor of $i$. Further to keep the representation of dynamical KMS symmetry manifest require the accompanying transformation~\cite{Jain:2020vgc} 
\be\label{ind}
     v^i \to v^i + \beta_0^{-1}(f_a^i-\Th \tilde f_a^i). 
\ee
The equations of motion for $X_a^i$ yield 
\be\label{nonc}
     \p_\mu T^{\mu i} = \Gamma^i,\quad \Gamma^i = \Lambda^{-1}_{ij} v^j + \cdots
\ee
where $T^{\mu i}$ is furnished by the Stückelberg action (see below) and $\Gamma^i$ by the non-Stückelberg action. Notice that redefinitions of $X_a^i$ and the corresponding induced redefinitions~\eqref{ind} have just enough degrees of freedom to ensure that $\Gamma^i$ has no higher-order corrections.  Thus, we may use field redefinitions to remove all higher-order corrections to $\sL^\text{(non)}$. These redefinitions cannot be used to remove any terms in $T^{\mu i}$ as $\sL^\text{(non)}$ is small even when it is lower-order in the derivative expansion. Further, we have no left-over field redefinitions to apply to the Stückelberg sector. As a result, $\sL^\text{(Stüc)}$ cannot be modified by field redefinitions of $X_a^i,v^i$. 

Now let us construct $\sL^\text{(Stüc)}$. Turning off external sources, the covariant building-blocks~are 
\bega\label{bb1}
     G_{ati} = \p_t X_{ai},\quad G_{aij} = \p_i X_{aj} + \p_j X_{ai},\\ 
     \vartheta_a = \theta_a - \ha \pp\times\vec X_a,\quad \hat \vartheta_a = \vartheta_a + i \beta_0  D_t \vartheta, \\
     \nabla_1 \theta = \vec n\cdot\pp\theta, \quad \nabla_1\theta_a = - \theta_a \vec n\times \pp \theta + \vec n \cdot \pp \theta_a - \vec n\cdot\pp X_a^i \p_i\theta , \\ \label{bb4}
     \nabla_2 \theta = \vec n\times\pp\theta,\quad \nabla_2 \theta_a = \theta_a \vec n\cdot\pp\theta - \vec n \times\pp \theta_a + \vec n\times \pp X_a^i \p_i\theta,
\end{gather}
where $ D_t \vartheta \equiv D_t\theta - \ha \omega$, with $\omega = \pp\times \vec v\,$ the pseudo-scalar vorticity. These building-blocks can be identified using the coset construction~\cite{Landry:2019iel,Landry:2021kko,Landry:2020obv}. We can now construct the Stückelberg sector Lagrangian
\be\begin{split}\label{pasL}
     \sL^\text{(Stüc)} = {i \gamma \ov \beta_0} \vartheta_a \hat \vartheta_a - K_1 \nabla_1 \theta \nabla_1\theta_a - K_2 \nabla_2 \theta \nabla_2\theta_a  \\ + \ha T_0^{ij} G_{aij} + T^{ti} G_{ati} + {i\ov 4\beta_0 } W^{ijkl} G_{aij} \hat G_{akl} , 
\end{split}\ee
for $T_0^{ij}=  p_0 \delta^{ij} + w_0 v^i v^j + \ha w_0 v^2 \delta^{ij}$, $T^{ti} = \tilde w_0 v^i$, and 
\be\label{W0}
     \begin{split}
          W^{ijkl} =   2\eta_\perp \Delta^{i (k}\Delta^{l)j} +\bigg(\zeta_\perp-\frac{2}{d}\eta_\perp \bigg) \Delta^{ij} \Delta^{kl} + 4 \eta_\parallel n^{(i} \Delta^{j)(k} n^{l)} \\ +\bigg(\zeta_\times -\frac{2}{d} \eta_\parallel\bigg) (\Delta^{ij} n^k n^l +\Delta^{kl} n^{i}n^{j} ) +\zeta_\parallel n^i n^j n^k n^l ,
     \end{split}
\ee
where $\Delta^{ij}= \delta^{ij} - n^i n^j$. The coefficients $\gamma , K_{1,2} , p_0,w_0,\tilde w_0,\eta_{\parallel,\perp},\zeta_{\parallel,\perp,\times}$  are all constant. Physically, $\eta_{\perp,\parallel}$ are shear viscosities, and $\zeta_{\perp,\parallel,\times}$ are bulk viscosities, $K_{1,2}$ are the Frank constants, and $p_0$ is the pressure. With boost symmetry restored, $w_0,\tilde w_0$ would coincide and equal the enthalpy density. The unitarity constraint~\eqref{unitary2} give positivity conditions on the dissipative transport coefficients, namely
\be\label{pos1}
     \gamma > 0,\quad \Lambda_0>0,\quad \Lambda_Q^2 < \Lambda_0^2 ,\quad \zeta_{\parallel,\perp} > 0,\quad \eta_{\parallel,\perp} > 0,\quad \zeta_\parallel \zeta_\perp > \zeta_\times^2 ,
\ee
and requiring the stability of the free energy implies $K_{1,2}\geq 0. $

The equation of motion for $\theta_a$ is
\be\label{th eom}
     \gamma D_t \vartheta = (K_1-K_2) \nabla_1\theta \nabla_2\theta + K_1 \pp\cdot(\vec n \, \nabla_1 \theta) - K_2\pp\times(\vec n \, \nabla_2 \theta),
\ee
which is simply~\eqref{eom a} modified to include a velocity. Next,
the equations of motion for $X_a^i$ give the non-conservation equation~\eqref{nonc} where $\Gamma^i = \Lambda^{-1}_{ij} v^j$, $T^{ti} = \tilde w_0 v^i$ and 
\be
     T^{ij} = p_0 \delta^{ij} + w_0 v^i v^j + \ha w_0 v^2 \delta^{ij} - K_1 \nabla_1 \theta n^{(i} \p^{j)} - K_2 \nabla_2 \theta n^k \epsilon^{k(i} \p^{j)} \theta - W_0^{ijkl} \p_k v_l. 
\ee
In the linearized limit, $v^i$ exponentially decays to zero. In the deep infrared, the leading-order equation of motion simply fixes $v^i=0$. Similarly the leading-order equation of motion for $\sigma^i$ is $X_a^i=0$. Plugging these back into the EFT yields~\eqref{probe}.

\section{Active dynamics with fuel}\label{active dyn}

To introduce activity to the nematic system described in the previous section, we simply couple it to a far-from-equilibrium sector. This sector will be described by a non-conserved $U(1)$ charge that begins at a large value and slowly decays, which models the burning of fuel. Active dynamics will take place on time scales much longer than the collision time, but much shorter than the decay time for the fuel. To ensure that such an intermediate regime exists and is large, we suppose that the fuel is described by approximate Stückelberg fields~$\varphi,\varphi_a$. 

We follow the formalism of~\cite{Landry:2020tbh}. If the fuel were exactly conserved, then we would introduce background $U(1)$ gauge sources $a_\mu,a_{a\mu}$ such that $\varphi,\varphi_a$ would always appear in the packages
\be
     A_\mu = a_\mu + \p_\mu \varphi,\quad A_{a\mu} = a_{a\mu} + \sL_{X_a} a_\mu + \p_\mu \varphi. 
\ee
To ensure that this $U(1)$ charge exists in normal (i.e. unbroken) phase, impose the time-independent shift symmetries~\cite{Dubovsky_2012}
\be\label{chem}
     \varphi \to \varphi + g(\sigma^i). 
\ee
It is convenient to define the local chemical potential\footnote{Typically a chemical potential is only well-defined when a charge is exactly conserved. Here because the non-conservation is very weak, an approximate notion of chemical potential can make sense on sufficiently short time scales. But eventually it will decay to zero, which is consistent with the common lore that chemical potentials for non-conserved quantities must vanish in equilibrium.}
\be
     \mu = D_t \varphi,\quad D_t = \p_t + \vec v\cdot \pp,
\ee
which is manifestly invariant under~\eqref{chem}.

As this $U(1)$ charge is not actually conserved, and $\varphi,\varphi_a$ are only approximate Stückelberg symmetries, we can decompose the Lagrangian by~\eqref{non stuc}. Like before, $\sL^\text{(non)}$ is considered~small. 

Dynamical KMS symmetry acts by~\eqref{dKMS1} and
\be
     \tilde \varphi = \Th \varphi,\quad \tilde \varphi_a = \Th \varphi_a + i \beta_0 \Th \mu. 
\ee
We want the chemical potential to begin at a rather large value, so we suppose that $\mu$ has a weight of zero, meaning that $\varphi_a$ also has a weight of zero. 

Let us begin by constructing the non-Stückelberg Lagrangian. It is given by
\be
     \sL^\text{(non)} = {i b\ov \beta_0} \varphi_a\hat\varphi_a + {i\Lambda^{-1}_{ij}\ov\beta} X_a^i \hat X_a^j,\quad \Lambda^{ij} = \Lambda_0 \delta^{ij} + \Lambda_Q Q^{ij} ,
\ee
where $\hat\varphi_a = \varphi_a + i \beta_0 \mu$. Here $b,\Lambda_{0,Q}$ are functions of $\mu$.\footnote{In principle they could also depend on $\varphi_a$ but we will truncate after quadratic order in $a$-type fields for simplicity.} Like the non-Stückelberg action discussed in the previous section, field redefinitions can be used to ensure that there are no higher-order corrections to this sector. These field redefinitions cannot be used to remove any terms from $\sL^\text{(Stüc)}$ as $\sL^\text{(non)}$ is considered small. Further, all field redefinitions have been used up, so terms of $\sL^\text{(Stüc)}$ that under ordinary circumstances could be removed via field redefinition, can no longer be removed. This fact will have important consequences. 

Let us now turn attention to the Stückelberg terms. When external sources are turned off, in addition to the building-blocks~\eqref{bb1}--\eqref{bb4}, we have 
\bega
     \mu,\quad \sA_a = D_t \varphi_a ,\quad A_{ai} ,\quad \hat \sA_a = \sA_a  + i \beta_0 D_t \mu,\quad  \hat A_{ai} = A_{ai} + i \beta_0 \p_i \mu.
\end{gather}
The Stückelberg Lagrangian up to weight-4 terms is
\be\begin{split}
     \sL^\text{(Stüc)} = {i \gamma \ov \beta_0} \vartheta_a \hat \vartheta_a - K_1 \nabla_1 \theta \nabla_1\theta_a - K_2 \nabla_2 \theta \nabla_2\theta_a  + T^{ti} G_{ati}+\ha T_0^{ij} G_{aij}  
     + n_0 \sA_a \\+ {i \sigma^{ij}\ov \beta_0} A_{ai} \hat A_{aj} + {i \kappa^{ij} \ov 2 \beta_0}(G_{aij} \hat \sA_{a} + \hat G_{aij}  \sA_{a}) 
     + {i\ov 4\beta_0 } W^{ijkl} G_{aij} \hat G_{akl} ,
\end{split}\ee
where $ T_0^{ij} = p_0 \delta^{ij} + w_0 v^i v^j + \ha w_0 v^2 \delta^{ij}$, $T^{ti} =\tilde w_0 v^i$, such that
\be
     w_0-\tilde w_0 = {\rm const.}, \quad n_0 = - {\p \sF \ov \p \mu } ,
\ee
for free energy\footnote{Note that free energy is a sensible quantity as long as some notion of local equilibrium, which is guaranteed by dynamical KMS symmetry, exists. }
\be
     \sF = -p_0(\mu) - \ha w_0 (\mu) v^2 + \ha K_1(\mu) (\nabla_1\theta)^2 + \ha K_2(\mu) (\nabla_2\theta)^2 . 
\ee
Stability of the free energy requires $K_{1,2} \geq 0$. 
All coefficients are functions of $\mu$ and we can decompose various tensor quantities by
\bega\label{0Q}
     \sigma^{ij} = \sigma_0 \delta^{ij} + \sigma_Q Q^{ij}  ,\quad \kappa^{ij} = \kappa_0 \delta^{ij} + \kappa_Q Q^{ij} ,
\end{gather}
and $W_0^{ijkl}$ is given by~\eqref{W0}. The unitarity constraint~\eqref{unitary2} places positivity conditions on dissipative transport coefficients~\eqref{pos1} and
\be\label{pos2}
     \sigma_0>0,\quad \kappa_0 > 0,\quad \sigma_Q^2 < 4 \sigma_0^2 ,\quad \kappa_Q^2 < 4\kappa_0^2.
\ee
Physically, $\sigma_{0,Q}$ are conductivities, $\eta_{\perp,\parallel}$ are shear viscosities, and $\zeta_{\perp,\parallel,\times}$ are bulk viscosities. The coefficients $\kappa_{0,Q}$ modify both the stress and the local fuel density $n_0$. Notice that if the $U(1)$ charge were conserved, we could work in Landau frame, thereby removing such terms. However, we already used up all field redefinitions for $\varphi,\varphi_a$ and $\sigma^i,X_a^i$ in the non-Stückelberg sector, making Landau frame impossible.\footnote{The fact that Landau frame is impossible is just an artefact of a choice we made. We found it convenient to make $\sL^\text{(non)}$ as simple as possible. But we could equally well have imposed Landau frame on $\sL^\text{(Stüc)}$ at the price of generating additional terms in $\sL^\text{(non)}$. } As a result, this term is in fact physical. This point is crucial as the $\kappa_Q$-term modifies the stress by an amount proportional to $Q^{ij}$, which is the hallmark of activity in nematics~\cite{Doostmohammadi1}.

Let us now investigate how this action gives rise to active dynamics. We will consider the equations of motion in the ``slow roll'' regime in which $\mu$ begins with a large and spatially homogeneous value and decays slowly. We further suppose that gradients in $\mu$ remain negligible. In this regime, the equations of motion for $\varphi_a$ are
\be
     \p_t \mu = - {\mu \ov\tau}  +\cdots ,\quad \tau = {\chi\ov b},\quad \chi \equiv {\p n_0 \ov \p \mu},
\ee
where $\cdots$ represent higher-order corrections. We therefore see that $\mu$ has instantaneous decay rate $\tau$. As $\sL^\text{(non)}$ is small, we must take $b$ small, meaning that $\tau$ is large, as expected. Next, the equations of motion for $X_a^i$ give the non-conservation equation~\eqref{nonc} where $\Gamma^i = \Lambda^{-1}_{ij} v^j$, $T^{ti} = \tilde w_0 v^i$ and 
\be
     T^{ij} = p_0 \delta^{ij} + w_0 v^i v^j + \ha w_0 v^2 \delta^{ij} - K_1 \nabla_1 \theta \, n^{(i} \p^{j)} - K_2 \nabla_2 \theta \, n^k \epsilon^{k(i} \p^{j)} \theta - W_0^{ijkl} \p_k v_l - \kappa^{ij} \p_t \mu. 
\ee
Expanding $\kappa^{ij}$ according to~\eqref{0Q} we find that the $\kappa_Q$-term augments the stress tensor with a term proportional to $Q^{ij}$, which is the hallmark of activity in nematic systems. 
Solving for $v^i$, we find that 
\be\label{act v}
     v^i =  (\alpha_0 \delta^{ij} + \alpha_Q Q^{ij} ) \p_k Q^{jk},\quad \alpha_{0,Q} \equiv  {\mu  \kappa_Q \Lambda_{0,Q} \ov \tau},
\ee
where we have dropped higher-order terms. The equations of motion for $\theta_a$ are, in the slow-roll approximation, given by~\eqref{th eom}. In conjunction with the active constitutive relation for the velocity~\eqref{act v}, we can compare with the  literature. We find that our expression for the velocity contains the standard activity $\alpha_0$ and also accounts for the anisotropic friction contribution to the activity $\alpha_Q$, proposed in~\cite{kawaguchi2017topological}. We emphasize that both of these terms contribute at the same order in the derivative expansion. As a result, unless there is fine-tuning, we cannot neglect either of these active terms. Notice that we must require $\tau$ to be large for there to be a sufficiently large separation of scales for interesting active dynamics to take place. So we might expect the activity $\alpha_{0,Q}$ must be small. Fortunately, this is not necessary as we may start with an arbitrarily large value of $\mu$. This is to say that the total amount of energy consumed can be very large even when the percent of fuel consumed per unit time is small so long as the total amount of fuel is large.

%\ML{Entropy current}

\section{Modified dynamical KMS symmetry}\label{mod dKMS}

Taking inspiration from the action in the previous section, we now attempt to modify dynamical KMS symmetry to describe active systems in the slow-roll regime on time-scales over which $\mu$ and $\p_t \mu$ do not change appreciably. To do this, remove $\varphi,\varphi_a$ from the EFT and introduce non-dynamical sources $M,M_a$ that transform under dynamical KMS symmetry by
\be
     \tilde M = \Th M,\quad \tilde M_a = \Th M_a + i \beta_0 \Th M. 
\ee
Notice that there is no time-derivative acting on $M$ in the expression for $\tilde M_a$. As a result, we choose $M,M_a$ to have opposite transformation properties under $\Th$. In particular, we take $\Th M_a = M_a(-t,\eta \vec x)$ and $\Th M = - M(-t,\eta \vec x)$, where $\eta=+1$ when $\Th$ contains no factors of $\sP$ and $\eta=-1$ when $\Th$ contains a factor of $\sP$. It is convenient to define $\hat M_a = M_a + i \beta_0 M$. 

The dynamical fields are now $X_a^i,\sigma^i$ and $\theta_a,\theta$. As $X_a^i,\sigma^i$ are still only approximate Stückelberg fields, we will continue to decompose the action according to~\eqref{non stuc}. The non-Stückelberg terms are
\be
     \sL^\text{(non)} = {i \Lambda^{-1}_{ij}\ov\beta} X_a^i \hat X_a^j,\quad \Lambda^{ij} = \Lambda_0 \delta^{ij} + \Lambda_Q Q^{ij} ,
\ee
for constants $\Lambda_{0,Q}$. 
And the Stückelberg terms are 
\be\begin{split}
     \sL^\text{(Stüc)} = {i \gamma \ov \beta_0} \vartheta_a \hat \vartheta_a - K_1 \nabla_1 \theta \nabla_1\theta_a - K_2 \nabla_2 \theta \nabla_2\theta_a 
     + T^{ti} G_{ati} +\ha T_0^{ij} G_{aij}  \\
     + {i \kappa^{ij} \ov 2 \beta_0}(G_{aij} \hat M_{a} + \hat G_{aij}  M_{a}) 
     + {i\ov 4\beta_0 } W^{ijkl} G_{aij} \hat G_{akl} .
\end{split}\ee
where~$W_0^{ijkl}$ is given by~\eqref{W0}, $T_0^{ij}=  p_0 \delta^{ij} + w_0 v^i v^j + \ha w_0 v^2 \delta^{ij}$, $T^{ti} = \tilde w_0 v^i$, and $\kappa^{ij} = \kappa_0\delta^{ij} + \kappa_Q Q^{ij}$ with constants $\gamma , K_{1,2} , p_0,w_0,\tilde w_0, \kappa_{0,Q},\eta_{\parallel,\perp},\zeta_{\parallel,\perp,\times}$. Positivity constraints from~\eqref{unitary2} are given by~\eqref{pos1} and~\eqref{pos2}. And requiring the free energy to be stable imposes $K_{1,2}\geq 0$. 

Turning off the $a$-type source $M_a=0$, setting $M$ constant, and solving the $X_a^i$-equations of motion for $v^i$, we find 
\be\label{actv}
     v^i =  (\alpha_0 \delta^{ij} + \alpha_Q Q^{ij} ) \p_k Q^{jk},\quad \alpha_{0,Q} \equiv  M  \kappa_Q \Lambda_{0,Q} ,
\ee
where sub-leading terms have been dropped. 
We therefore see that $M$ is the external source for activity; taking $M=0$ recovers the passive Lagrangian~\eqref{pasL}. The equations of motion for $\theta_a$ are given by~\eqref{th eom}, which in conjunction with~\eqref{actv}, yields active dynamics. Notice that the $\alpha_Q$-term accounts for the effects of anisotropic friction~\cite{kawaguchi2017topological}. 

An important note is that dynamical KMS symmetry ensures the second law of thermodynamics. That is, it provides a prescription for constructing an entropy current $s^\mu$ whose on-shell divergence is non-negative, $\p_\mu s^\mu \geq 0$. This is, however, only guaranteed when all external sources are turned off, $M=M_a=0$. As a result, the Lagrangian constructed in this section will not satisfy the second law of thermodynamics. This finding should not be surprising: by replacing the $U(1)$ non-conserved fuel charge with external sources, we are no longer able to keep track of the entropy produced by burning fuel.

Notice that the EFT constructed in the previous section is more complete in that it gives a detailed account of how activity is generated. While this may be good for some purposes, it also has the downside of being model-dependent. By contrast the EFT constructed here is agnostic about the mechanism of activity and is therefore model-independent. All predictions of this theory are, however, only valid up to corrections of order $O(1/\tau)$, where $\tau$ is the characteristic decay time of the fuel.

Finally, it is important to note that setting $M$ constant and $M_a=0$ is not the only possibility for generating active dynamics. This is the correct prescription if our source of activity is a slowly decaying fuel, but in principle, activity could be generated from an inherently noisy process. In such cases, if the noise profile of the activity is known, then $M,M_a$ can be treated as stochastic variables. This stochastic behavior could lead to different relations between the equations of motion and noise profiles of the active system. In this way, such a theory will give an unambiguous generalization of the fluctuation-dissipation~relations.

%\ML{What is the new fluctuation-dissipation theorem?}

\section{Discussion}\label{discussion}

In this work, we constructed non-equilibrium EFTs defined on the SK contour for active nematics. We took two approaches, one that describes a particular mechanism for activity and the other, which is more model-independent.\footnote{It is not absolutely model-independent as we still suppose that {\it some} notion of local thermal equilibrium exists. To account for the case in which no notion of local equilibrium exists in trivial: simply discard dynamical KMS symmetry. } 
%We found leading-order corrections to the standard equations of motion typically found in the literature. In particular, the active constitutive relation for the velocity that we found has two independent terms
%\be
%     v^i = (\alpha_0\delta^{ij} + \alpha_Q Q^{ij})\p_k Q^{jk},\quad \alpha_Q^2 \leq 4 \alpha_0^2 . 
%\ee
%Notice that the above expression is not the most general expression for $v^i$ in terms of $Q^{ij}$ at one-derivative order (e.g. a term proportional to $\epsilon^{ij}\p_j \theta$ might appear). The fact that additional terms do not appear is a non-trivial constraint imposed by dynamical KMS symmetry. Further, 
Our EFTs are based entirely on symmetry principles organized in a derivative expansion that account for both classical equations of motion and statistical fluctuations about the mean. As a result, they give unambiguous predictions regarding the relationship between statistical fluctuations and classical equations of motion. That is, these EFTs generalize the famous fluctuation-dissipation relations to a wide class of active system. A potentially fruitful further direction is to study defects~\cite{vafa2020multi-defect,vafa2022defectAbsorption,vafa2022defectDynamics,vafa2023active,vafa2022active} using EFT. In particular, the language of field theory is well-suited to studying phase transitions, so EFT may shed light on the activity-induced BKT-type phase transition~\cite{PhysRevLett.121.108002}.

While we focused on constructing EFTs for active nematics, the principles exploited in this paper can be easily generalized to active systems more broadly. We propose the following prescription for constructing model-independent active actions:
\begin{enumerate}
     \item Identify the degrees of freedom associated with the passive system in the probe limit. 
     \item Introduce non-Stückelberg fields $\sigma^i,X_a^i$, which describe velocity. Treat them as approximate Stückelberg fields and decompose the Lagrangain by
     \be
          \sL = \sL^\text{(non)} + \sL^\text{(Stüc)},
     \ee
     where $\sL^\text{(non)}$ is small and contains non-Stückelberg terms, while $\sL^\text{(Stüc)}$ contains Stückelberg terms. Suppose all fields transform under dynamical KMS symmetry in the usual way. 
     \item Couple the action to active sources $M,M_a$ that transform under the dynamical KMS symmetry by
     \be
          \tilde M = \Th M,\quad \tilde M_a = \Th M_a + i\beta_0 \Th M,
     \ee
     where $M,M_a$ transform with opposite signs under $\Th$.
     \item Identify a derivative power-counting scheme and construct the most general action consistent with symmetries up to a given order. 
\end{enumerate}
For specific models of activity, instead of introducing sources $M,M_a$, one could couple the system to a far-from-equilibrium sector, like the non-conserved $U(1)$ fuel charge discussed in this work.

\bigskip

\bigskip

\noindent {\bf ACKNOWLEDGMENTS:}  I would like to thank Farzan Vafa and Matteo Baggioli for many insightful discussions and comments on the manuscript. This work was supported by the ALFA Foundation.

\bibliographystyle{apsrev4-1}
\bibliography{refs}{}

\end{document}